\documentclass[pdflatex,sn-mathphys-num]{sn-jnl}

\usepackage{graphicx}%
\usepackage{multirow}%
\usepackage{amsmath,amssymb,amsfonts}%
\usepackage{amsthm}%
\usepackage{mathrsfs}%
\usepackage[title]{appendix}%
\usepackage{xcolor}%
\usepackage{textcomp}%
\usepackage{manyfoot}%
\usepackage{booktabs}%
\usepackage{algorithm}%
\usepackage{algorithmicx}%
\usepackage{algpseudocode}%
\usepackage{listings}%

\theoremstyle{thmstyleone}%
\theoremstyle{thmstyletwo}%

\theoremstyle{thmstylethree}%

\begin{document}

\title[Article Title]{A Robust Compressible APIC–FLIP Particle–Grid Method with Conservative Resampling and Adaptive APIC–PIC Blending}


\author*{\fnm{Jiansheng} \sur{Yao}}\email{ab135794@ustc.edu.cn}

\author*{\fnm{Yingkui} \sur{Zhao}}\email{Zhao\_yingkui@iapcm.ac.cn}

\affil{\orgname{Institute of Applied Physics and Computational Mathematics}, \orgaddress{\city{Beijing}, \postcode{100084},\country{China}}}


\abstract{Modeling inviscid compressible flows with shocks and vortex-dominated dynamics remains challenging for particle–grid
methods due to moving discontinuities, cell-crossing noise, and quadrature degradation under strong deformation.
Building on a FLIP–APIC framework with vorticity-aware tensor artificial viscosity, we identify a long-time RTI
failure mode: particle depletion at spike heads degrades quadrature and particle–grid coupling, producing
nonphysical, void-like dents.

Standard mitigations (CPDI-lite and subcell-jittered seeding) reduce but do not eliminate this artifact.
We therefore add two sampling-aware controls: (i) conservative split resampling that replenishes depleted cells
while exactly conserving mass, momentum, and internal energy; and (ii) a soft-switch that attenuates only the APIC
affine term when local support is insufficient. Tests on the Sod shock tube and single-/multi-mode RTI show that
the method removes spike-head voids in long-time RTI while preserving vortex roll-up, and matches reference Euler
growth metrics.
}




\maketitle

\section{Introduction}\label{sec1}

Robust discontinuity control and shock capturing are central requirements in compressible-flow simulation,
particularly for problems involving shock propagation, multi-material interfaces, and vortical dynamics.
Conventional Eulerian schemes have achieved great success; however, fixed spatial discretizations may diffuse
shocks and interfaces unless high-resolution shock-capturing techniques (e.g., Riemann solvers, limiters, or
adaptive refinement) are employed, and convection-dominated regimes typically rely on upwinding for stability.~\cite{WoodwardColella1984,Roe1986,ShuOsher1988}
In contrast, Lagrangian descriptions follow fluid parcels, reducing convective transport error and helping to
preserve material interfaces, although additional stabilization remains necessary in the presence of shocks and
moving discontinuities. In this context, particle-based approaches are attractive because they avoid severe mesh
distortion and remeshing. Among them, the material point method (MPM) combines Lagrangian material points with a
background grid: particles act as moving quadrature points, while the grid provides a convenient numerical
workspace for force evaluation and updates.~\cite{Sulsky1994MPM,deVaucorbeil2020MPMReview}

MPM is closely related to classical particle-in-cell (PIC) schemes~\cite{Harlow1964PIC}, and inherits low-dissipation transfer variants such as FLIP~\cite{BrackbillRuppel1986FLIP,BurgessSulskyBrackbill1992FLIPMassMatrix}, weighted least-squares PIC transfers~\cite{WallstedtGuilkey2011WLS}, and APIC~\cite{Jiang2015APIC}. To mitigate cell-crossing noise and quadrature degradation under large deformation, widely used extensions include GIMP~\cite{BardenhagenKober2004GIMP} and CPDI/CPDI2~\cite{Sadeghirad2011CPDI,Sadeghirad2013CPDI2}, as well as improved particle updates and accuracy analyses~\cite{Steffen2008Quadrature,HammerquistNairn2017Noise,NairnHammerquist2021MassMatrixInv,Stefan2023StrongShock}. Reproducing-kernel variants further enhance consistency in particle--grid coupling~\cite{Liu1995RKPM,Huang2019RKPM2D}. Canonical strong-shock benchmarks for validating compressible formulations include the Sod shock tube and the Sedov blast wave~\cite{Sod1978,Sedov1993Similarity}.

Despite these advantages, compressible MPM must balance robustness near shocks with fidelity in vortical and
shear-dominated regions. Moving discontinuities and shock-driven transients can trigger oscillations unless
sufficient numerical dissipation is introduced. Artificial viscosity (AV) is therefore widely used to spread
shocks over a few grid cells, damp spurious oscillations, and improve robustness; the quadratic AV term controls
shock thickness through its nonlinear dependence on velocity divergence. However, classical von
Neumann--Richtmyer--Landshoff AV exposes a trade-off between stability and accuracy in multi-dimensional flows,~\cite{VonNeumannRichtmyer1950,Landshoff1955,Wilkins1980,Noh1987,KolevRieben2009}
especially when shocks coexist with shear layers and vortical roll-up. Peddavarapu and Huang (2025)~\cite{PeddavarapuHuang2025} identify two
key limitations: (i) a lack of deviatoric/shear contributions and (ii) an overly restrictive linear term in
vortical regimes. They propose a vorticity-aware tensor-viscosity formulation that improves stability while
better preserving vortical structures.

Building on this line of work, we target a complementary but practically critical failure mode in compressible
particle--grid solvers: \emph{under-sampling} (quadrature deficiency) in strongly deforming regions. This
pathology is prominent in long-time Rayleigh--Taylor instability (RTI)~\cite{Taylor1950RTI,Kull1991RTITheory}, where the spike head can become
transiently particle-depleted under severe stretching. The resulting loss of particle support degrades local
quadrature and particle--grid coupling, manifesting as nonphysical pressure/density depressions and, in severe
cases, pit-like voids at the spike head. Moreover, while higher-order particle--grid transfers such as APIC are
desirable for resolving vortex-rich shear layers, poorly supported affine modes can become ill-conditioned in
depleted cells and inject spurious energy, further amplifying spike-head artifacts.

To address these coupled issues, we propose a robust compressible particle--grid framework that extends the
FLIP--APIC + vorticity-aware tensor-AV baseline of Peddavarapu and Huang (2025) with sampling-aware controls.
We first incorporate two standard mitigations against sampling bias: CPDI-lite transfers to reduce quadrature
error under imperfect particle distributions~\cite{Sadeghirad2011CPDI}, and a small subcell jitter in the initial particle seeding to
break phase locking and suppress grid imprinting~\cite{Steffen2008Quadrature,HammerquistNairn2017Noise}. While these measures substantially reduce the artifact, small
residual dents may still appear in long-time RTI runs when the spike head becomes transiently depleted; once
triggered, such dents can evolve non-physically and deepen into pit-like voids, compromising the late-time
morphology. We therefore introduce two additional remedies that directly address particle-support loss and affine
instability: (i) a conservative resampling strategy that splits particles in depleted cells while exactly
conserving mass, momentum, and internal energy, thereby restoring local quadrature quality; and (ii) a cell-wise
soft-switch that continuously down-weights only the APIC affine contribution toward PIC-like behavior when local
particle support falls below prescribed thresholds, preventing spurious affine energy injection precisely where
particle deficiency is most severe. Together, these components retain the shock/discontinuity robustness of tensor
AV and the strong interface-advection properties of FLIP--APIC in well-sampled regions, while eliminating
sampling-driven spike-head void formation and preserving physically meaningful vortex roll-up in challenging
long-time RTI regimes.

The paper is organized as follows. Section~2 reviews governing equations and notation. Section~3 summarizes the
baseline behavior and highlights the spike-head failure mode. Section~4 details the proposed transfers, viscosity
model, sampling-aware soft-switch, and conservative resampling. Section~5 presents validation and ablation
studies on standard benchmarks and single-/multi-mode RTI, followed by discussion and conclusions in
Section~6.

\section{Basic equations}\label{sec2}

\subsection{Governing equations}
We consider inviscid, compressible flow governed by the Euler equations in conservative form:
\begin{equation}
\frac{\partial \rho}{\partial t} + \nabla\cdot(\rho\mathbf{u}) = 0,
\label{eq:euler_mass}
\end{equation}
\begin{equation}
\frac{\partial (\rho\mathbf{u})}{\partial t} + \nabla\cdot\!\left(\rho\mathbf{u}\otimes\mathbf{u} + p\mathbf{I}\right) = \rho\mathbf{g},
\label{eq:euler_momentum}
\end{equation}
\begin{equation}
\frac{\partial (\rho E)}{\partial t} + \nabla\cdot\!\left((\rho E + p)\mathbf{u}\right) = \rho\,\mathbf{g}\cdot\mathbf{u}.
\label{eq:euler_energy}
\end{equation}
Here $\rho$ is density, $\mathbf{u}$ is velocity, $p$ is pressure, $\mathbf{I}$ is the identity tensor,
$\mathbf{g}$ is body acceleration, and $E=e+\tfrac{1}{2}\lVert \mathbf{u}\rVert^{2}$ is the total specific
energy with internal energy $e$. We close the system with the ideal-gas equation of state:
\begin{equation}
p = (\gamma-1)\rho e.
\label{eq:eos_ideal_gas}
\end{equation}
The local sound speed is $c=\sqrt{\gamma p/\rho}$.

We adopt a standard particle--grid splitting in which Lagrangian particles (material points) carry state
variables and act as moving quadrature points, while a background grid is used for spatial discretization
and time stepping.

Throughout the remainder of this paper, the subscript $p$ denotes particles (material points) and the
subscript $i$ denotes grid nodes. Each particle $p$ stores position $\mathbf{x}_p$, velocity $\mathbf{v}_p$,
mass $m_p$, specific internal energy $e_p$, density $\rho_p$, and volume $V_p$.
In APIC mode, each particle also stores an affine matrix (e.g., $\mathbf{C}_p$) representing a local
velocity-gradient proxy. Grid node $i$ stores lumped mass $m_i$ and velocity $\mathbf{v}_i$.
At each time step, we (i) reset grid fields, (ii) transfer particle mass and momentum to the grid (P2G),
(iii) apply body forces and internal forces (pressure and AV) to update grid velocities, (iv) transfer
updated grid velocities back to particles (G2P) using a PIC/FLIP/APIC blend, (v) advect particles, and
(vi) optionally apply conservative resampling every $N_{\mathrm{res}}$ steps (Algorithm~1 summarizes the
full procedure).

\subsection{Material point method basic equations}
We briefly summarize the standard MPM semi-discrete update used by particle--grid methods.
Let $N_i(\mathbf{x})$ denote the grid shape function associated with node $i$ and define
\begin{equation}
w_{ip}=N_i(\mathbf{x}_p), \qquad \nabla w_{ip}=\nabla N_i(\mathbf{x}_p).
\label{eq:shape_weights}
\end{equation}

\paragraph{Particle-to-grid (mass and momentum).}
Grid lumped mass and momentum are assembled by quadrature:
\begin{equation}
m_i = \sum_p w_{ip}\, m_p, \qquad
(m\mathbf{v})_i = \sum_p w_{ip}\, m_p\, \mathbf{v}_p ,
\label{eq:mpm_p2g_basic}
\end{equation}
leading to the nodal velocity $\mathbf{v}_i=(m\mathbf{v})_i/m_i$ after enforcing boundary conditions.
(Section~4 later extends \eqref{eq:mpm_p2g_basic} with FLIP/APIC transfers.)

\paragraph{Grid momentum equation and internal force.}
MPM advances momentum on the grid using a lumped-mass ODE:
\begin{equation}
m_i\,\frac{d\mathbf{v}_i}{dt} = \mathbf{f}_i^{\mathrm{int}} + \mathbf{f}_i^{\mathrm{ext}},
\label{eq:mpm_grid_momentum}
\end{equation}
where the body-force contribution can be written as
\begin{equation}
\mathbf{f}_i^{\mathrm{ext}} = \sum_p w_{ip}\, m_p\,\mathbf{g}.
\label{eq:mpm_external_force}
\end{equation}
The internal force is obtained from the particle Cauchy stress $\boldsymbol{\sigma}_p$:
\begin{equation}
\mathbf{f}_i^{\mathrm{int}} = -\sum_p V_p\, \boldsymbol{\sigma}_p\, \nabla w_{ip}.
\label{eq:mpm_internal_force}
\end{equation}
For inviscid compressible flow we use the pressure stress
$\boldsymbol{\sigma}_p^{(p)} = -p_p\,\mathbf{I}$, optionally augmented by an artificial-viscosity stress
$\boldsymbol{\sigma}_p^{(\mathrm{av})}$ (Section~4.3), i.e.,
$\boldsymbol{\sigma}_p = -p_p\mathbf{I} + \boldsymbol{\sigma}_p^{(\mathrm{av})}$.

With explicit time integration, the nodal velocity update reads
\begin{equation}
\mathbf{v}_i^{n+1} = \mathbf{v}_i^{n} + \Delta t\,
\frac{\mathbf{f}_i^{\mathrm{int}}+\mathbf{f}_i^{\mathrm{ext}}}{m_i}.
\label{eq:mpm_grid_update}
\end{equation}

\paragraph{Grid-to-particle (velocity update) and advection.}
After updating $\mathbf{v}_i^{n+1}$, particle velocities are reconstructed by interpolation (PIC form)
\begin{equation}
\mathbf{v}_p^{n+1} = \sum_i w_{ip}\,\mathbf{v}_i^{n+1},
\label{eq:mpm_g2p_pic}
\end{equation}
and particles are advected as
\begin{equation}
\mathbf{x}_p^{n+1} = \mathbf{x}_p^{n} + \Delta t\,\mathbf{v}_p^{n+1}.
\label{eq:mpm_advect}
\end{equation}
(Section~4.1 uses a PIC/FLIP blend and an APIC affine enhancement.)

\paragraph{Thermodynamics and volume update.}
To update compressible thermodynamic variables consistently, we use the particle-wise work rate
\begin{equation}
m_p\,\frac{de_p}{dt} = - V_p\, \boldsymbol{\sigma}_p : \nabla \mathbf{v}(\mathbf{x}_p),
\label{eq:mpm_energy_work}
\end{equation}
where $\nabla \mathbf{v}(\mathbf{x}_p)$ is reconstructed from the grid velocity field, e.g.,
\begin{equation}
\nabla \mathbf{v}(\mathbf{x}_p)=\sum_i \mathbf{v}_i \otimes \nabla w_{ip}.
\label{eq:mpm_vel_grad}
\end{equation}
The particle volume can be updated using $\dot V_p = V_p\, \nabla\cdot \mathbf{v}(\mathbf{x}_p)$, so that
$\rho_p=m_p/V_p$ remains consistent, and pressure is closed by the EOS \eqref{eq:eos_ideal_gas}.

\section{BASELINE RECAP AND FAILURE MODES }\label{sec3}

\subsection{Baseline performance on Richtmyer–Meshkov instability }
As a baseline, we reproduce the compressible-flow MPM variant combining FLIP–APIC transfers with a vorticity-aware tensor artificial viscosity (following Peddavarapu and Huang, 2025). This baseline captures Richtmyer–Meshkov (RM) instability robustly and serves as a reference configuration. A background grid with a nodal discretization of 76 × 101 is employed, and the foreground gas domain consists of 175,851 material points (351 × 501). Under these settings, the average initial PPC ratio in the foreground domain is roughly 25. Sliding boundary conditions are applied to all boundaries of the background grid. Figure 1 shows an RM result obtained with the tensor artificial viscosity under the same initial and boundary conditions as in Peddavarapu and Huang (2025); the overall morphology is qualitatively consistent with the fourth row of their Fig. 32. In compression-dominated regions the artificial viscosity suppresses numerical instabilities effectively, while in shear-dominated regions it reduces excessive diffusion and better preserves vortical roll-up.
\begin{figure}
    \centering
    \includegraphics[width=0.95\linewidth]{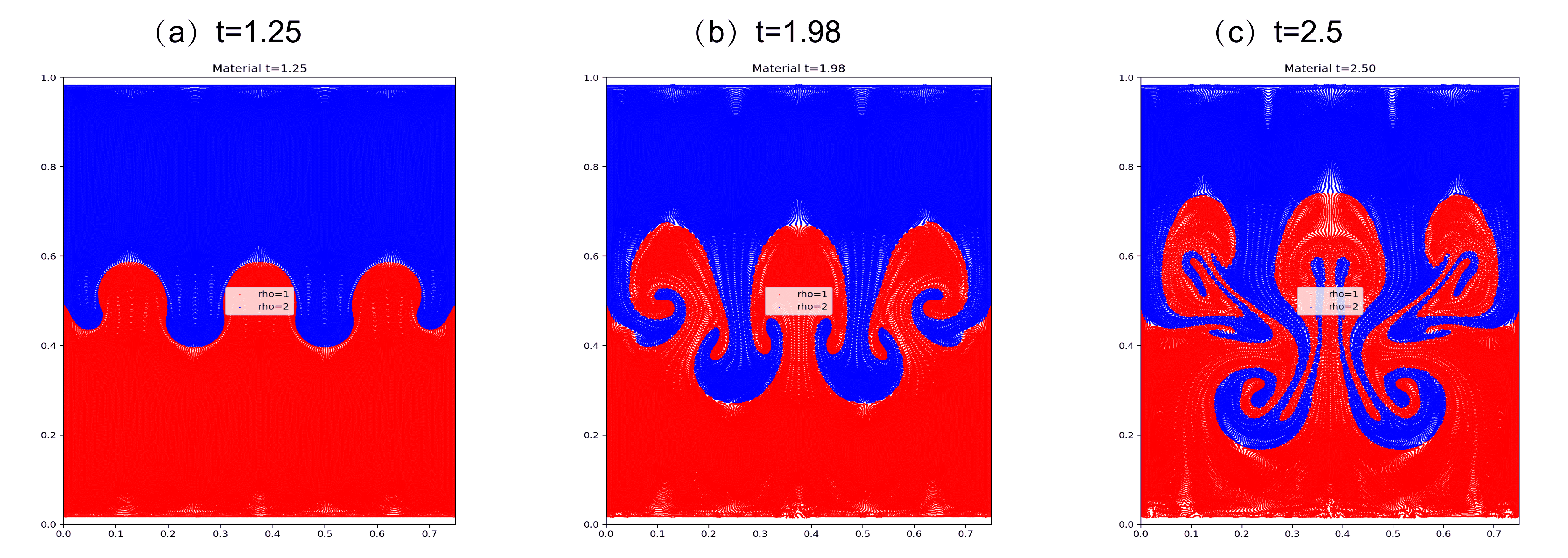}
    \caption{Richtmyer–Meshkov instability in the FLIP–APIC + vorticity-aware tensor-AV baseline. Particles are colored by material (red/blue), illustrating post-shock interfsace roll-up at different time.
}
    \label{fig:rmi_baseline}
\end{figure}
\begin{figure}
    \centering
    \includegraphics[width=0.95\linewidth]{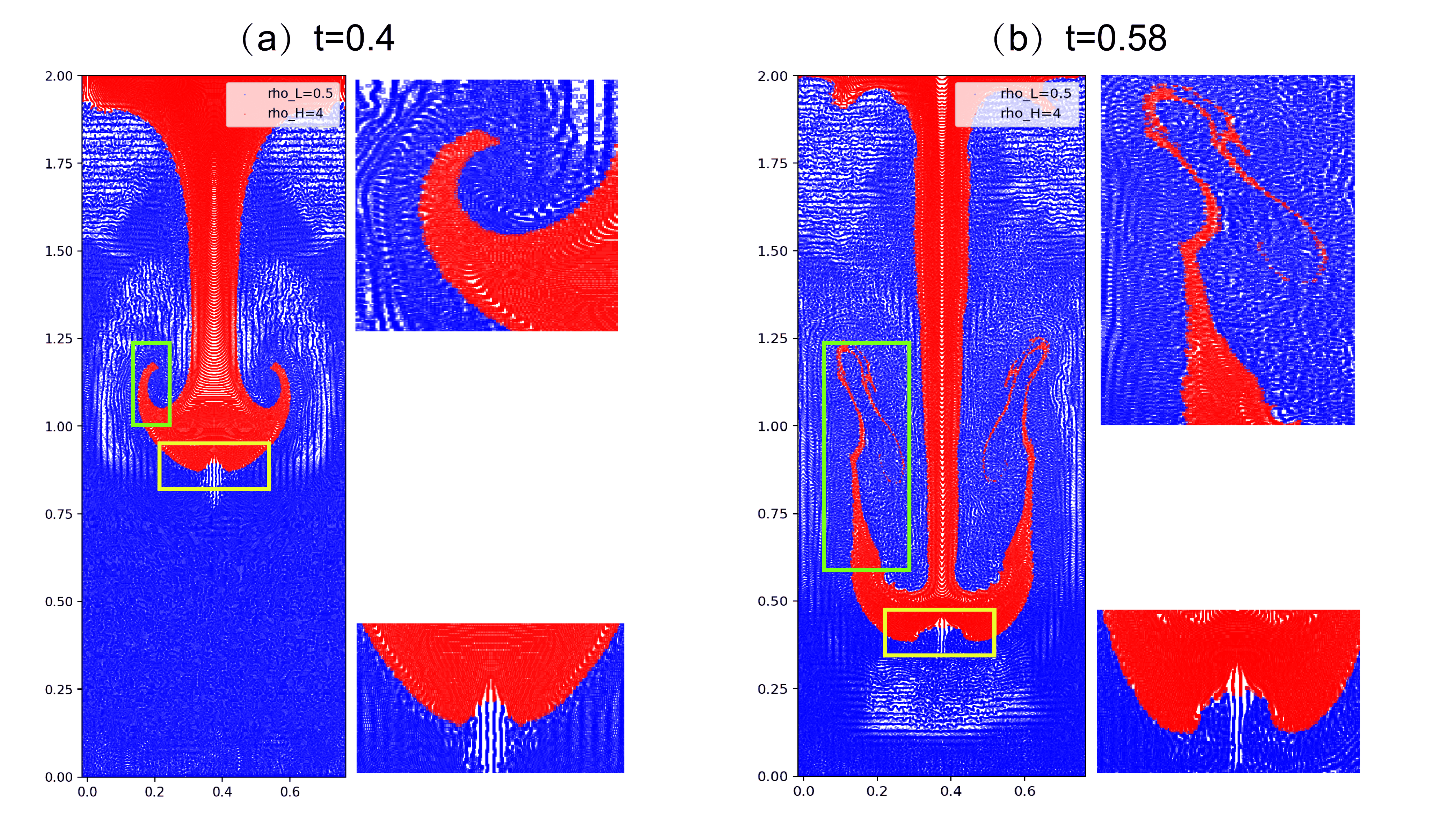}
    \caption{ Single-mode RTI in the FLIP–APIC + tensor-AV baseline: a nonphysical spike-head dent/void-like depression develops in long-time evolution (highlighted region).
}
    \label{fig:rti_dent_baseline}
\end{figure}
\subsection{Nonphysical spike-head dent in long-time RTI evolution }
In Fig.~1, the RM spikes and bubbles evolve only within $y\in[0,1]$ and quickly interact with the sliding
boundaries, which limits further spike/bubble development and restricts the accessible evolution time. We
therefore consider a long-time single-mode RTI test in an enlarged domain. Unless otherwise stated, we use a
nodal discretization of $76\times 201$ with $\mathrm{PPC}=16$, periodic boundaries in $x$, and reflecting
boundaries in $y$. The heavy and light fluids have densities $\rho_h=4$ and $\rho_\ell=0.5$, initially separated
by an interface at $y=1.5$. To accelerate the instability growth, we use an initial perturbation amplitude of
$0.1$ and set the gravitational acceleration to $g=15$.

Figure~2 shows that the baseline method can develop a spike-head artifact in this long-time RTI regime. At
$t=0.4$ (Fig.~2(a)), the spike head (yellow box) is already visibly under-sampled, and a small indentation is
observed at the tip. As the spike continues to stretch, local particle support further degrades; the depleted
region expands and the dent increases in extent and depth, becoming clearly pronounced by $t=0.58$ (Fig.~2(b)).
Once formed, the dent may persist and deepen over time, biasing the late-time spike-head morphology.

We attribute this behavior to sampling-driven quadrature degradation near the spike tip. As particles evacuate
cells in the spike-head region, the available particle support becomes insufficient for accurate integration of
pressure and viscous forces, which can bias the reconstructed force balance and induce an artificial local
under-pressure. This feedback can sustain the indentation and, in severe cases, evolve into a void-like
depression. These observations motivate the sampling-aware resampling and transfer-gating mechanisms introduced
in Sec.~\ref{sec:undersampling_controls}.

\section{Method}\label{sec4}

We build on the compressible FLIP--APIC particle--grid framework equipped with the vorticity-aware tensor
artificial viscosity (tensor AV) of Peddavarapu and Huang (2025). In this class of solvers, particles act as
moving quadrature points carrying mass, momentum, and thermodynamic state, while a background grid is used to
assemble internal forces (pressure and AV stresses) and advance nodal momentum. Particle velocities are updated
with an A-FLIP blend, and APIC further augments the particle state with an affine matrix $\mathbf{C}_p$ to
reduce grid-crossing artifacts and better preserve vortex-rich shear-layer dynamics.

Our focus is long-time Rayleigh--Taylor instability (RTI), where severe stretching near spike heads can
transiently evacuate particles from a small set of cells. This under-sampling leads to two coupled numerical
issues: (i) quadrature degradation, where insufficient particle support biases particle--grid transfers and the
reconstruction of density/pressure; and (ii) affine-mode instability, where the APIC affine matrix
$\mathbf{C}_p$ becomes poorly supported and its contribution can inject spurious kinetic energy in depleted
regions. In the baseline FLIP--APIC + tensor-AV solver, these effects can manifest as a persistent, nonphysical
spike-head dent at late times.

To reduce sampling bias for a given particle set, we adopt two standard mitigations throughout this work.
First, we employ a lightweight CPDI-like four-corner averaging (``CPDI-lite'') for particle--grid weights~\cite{Sadeghirad2011CPDI},
which locally anti-aliases the transfer and reduces grid imprinting/phase locking and quadrature bias under
imperfect particle distributions (Sec.~4.1, Eq.~(4)--(4a)). Second, we apply a small subcell jitter to the
initial particle seeding to break strict lattice alignment with the grid while preserving the macroscopic
initial condition~\cite{Steffen2008Quadrature,HammerquistNairn2017Noise} (Sec.~4.2, Eq.~(12)). These measures substantially weaken sampling-driven artifacts; however,
they cannot fully prevent late-time dents when the spike head becomes \emph{severely} depleted: CPDI-lite can
only smooth the contribution of \emph{existing} particles and cannot restore missing support, and $\mathbf{C}_p$
may still become ill-conditioned during brief depletion events.

We therefore introduce two additional under-sampling controls that directly target particle-support loss and
affine instability in depleted cells. The first is a conservative split/merge resampling strategy that
replenishes particles in depleted cells to restore local quadrature quality while exactly conserving mass,
momentum, and internal energy. The second is a smooth, cell-wise soft-switch that attenuates only the APIC
affine contribution toward PIC-like behavior when local support drops below prescribed thresholds, preventing
spurious affine-mode energy injection while retaining APIC accuracy in well-sampled shear layers
(Sec.~\ref{sec:undersampling_controls}). Together, these components preserve the strong shock/discontinuity
robustness of tensor AV and the low-dissipation advection of FLIP--APIC in well-sampled regions, while improving
long-time stability in sampling-limited RTI regimes.

For clarity, Algorithm~\ref{algo:apicflip} summarizes the complete time-stepping loop. The subsequent
subsections then provide the precise transfer definitions (including CPDI-lite), initialization strategy,
thermodynamics and time stepping, tensor AV, and the detailed design of the under-sampling controls.

\begin{algorithm}
\caption{APIC/FLIP time stepping with tensor artificial viscosity (AV), Biindicator-gated APIC affine transfer, and conservative resampling}\label{algo:apicflip}
\begin{algorithmic}[1]
\Require Initial particles $\{\mathbf{x}_p^{0},\mathbf{v}_p^{0}\}$ (or prescribed perturbation), phase/material map for $\rho_0(\mathbf{x}),p_0(\mathbf{x})$, grid spacing $\Delta x$, time step $\Delta t$, EOS parameter $\gamma$, AV coefficients $(c_l,c_q)$, A-FLIP mixing $\alpha$, resampling period $N_{\mathrm{res}}$.
\Ensure Updated particle states $\{\mathbf{x}_p^{n+1},\mathbf{v}_p^{n+1},V_p^{n+1},e_p^{n+1},\mathbf{C}_p^{n+1}\}$ and grid states at step $n+1$.

\State \textbf{(State definition):} Given $\{\mathbf{x}_p^{n},\mathbf{v}_p^{n},m_p,V_p^{n},e_p^{n},\mathbf{C}_p^{n}\}$; compute $\rho_p^{n}=m_p/V_p^{n}$ and $p_p^{n}=(\gamma-1)\rho_p^{n}e_p^{n}$ when needed.

\State \textbf{(Support statistics):} (Optional) compute per-cell particle counts $n_c$ for support gating and/or resampling decisions.

\State \textbf{(Biindicator gate):} (Optional) update $\omega_p\in[0,1]$ from smooth weights (support $w_{\mathrm{Sup}}$ and a combined compression--vorticity factor $w_{A\Omega}$), e.g.,
$\omega_p=\sqrt{w_{\mathrm{Sup}}\,w_{A\Omega}}$; if disabled set $\omega_p\gets 1$.

\State \textbf{(Grid reset):} reset grid masses $m_i \gets 0$, grid momenta $(m\mathbf{v})_i \gets \mathbf{0}$ (and any force/gradient accumulators).

\State \textbf{(P2G mass \& momentum):} for each particle $p$ and neighboring node $i$ with weight $w_{ip}$,
\[
m_i \gets m_i + w_{ip}m_p,\qquad
(m\mathbf{v})_i \gets (m\mathbf{v})_i + w_{ip}m_p\Big[\mathbf{v}_p + \omega_p\,\mathbf{C}_p(\mathbf{x}_i-\mathbf{x}_p)\Big].
\]

\State \textbf{(Velocity normalization \& BC):} compute grid velocities $\mathbf{v}_i \gets (m\mathbf{v})_i/m_i$ and apply boundary conditions.

\State \textbf{(Grid update with EOS + AV):} compute pressure by EOS and tensor AV; update grid velocities using body forces and internal forces.

\State \textbf{(G2P reconstruction):} reconstruct $\mathbf{v}_{p,\mathrm{PIC}}^{n+1}$ and $\Delta\mathbf{v}_{p,\mathrm{FLIP}}^{n+1}$ from the grid; update
$\mathbf{v}_p^{n+1}$ via A-FLIP blending (cf.\ Eq.~(6)).

\State \textbf{(Particle advection):} advect particles by
$\mathbf{x}_p^{n+1} \gets \mathbf{x}_p^{n} + \Delta t\,\mathbf{v}_p^{n+1}$.

\State \textbf{(Thermodynamics \& APIC affine):} update $V_p,e_p$ (hence $\rho_p$) using work terms and EOS; reconstruct $\mathbf{C}_p$ (APIC), apply clipping if desired, and store $\mathbf{C}_p \gets \omega_p\,\mathbf{C}_p$.

\State \textbf{(Conservative resampling):} every $N_{\mathrm{res}}$ steps, split particles in depleted cells and (optionally) merge particles in over-dense cells, conserving mass, momentum, and internal energy.

\end{algorithmic}
\end{algorithm}

We next detail each component of Algorithm~\ref{algo:apicflip}, starting with the particle--grid transfer
operators and the CPDI-lite weights used to reduce quadrature bias.

\subsection{Particle–grid transfers (PIC/FLIP/APIC)}

We use quadratic B-spline shape functions for particle–grid interpolation. Let $w_{ip}$ denote the weight between material point $p$ at position $\mathbf{x}_p$ and grid node $i$ at position $\mathbf{x}_i$. Grid nodal mass and momentum are assembled from the material points, and nodal velocities are obtained by mass lumping.

\paragraph{CPDI-lite transfer for reduced quadrature bias}
Point-sampled transfers can suffer from grid imprinting and phase locking when particles are seeded on a regular lattice, and they become particularly noisy when the local particle support deteriorates. To reduce these sampling-induced biases without introducing a full domain-tracking CPDI formulation~\cite{Sadeghirad2011CPDI}, we use a light-weight CPDI-like \emph{four-corner} sampling (``CPDI-lite'') in which the particle-to-grid weight is replaced by an average over a small subcell stencil around the particle:
\begin{equation}
w^{\mathrm{lin}}_{ip}=N_i(\mathbf{x}_p),
\qquad
w^{\mathrm{cpdi}}_{ip}=\frac{1}{4}\sum_{c=1}^{4} N_i(\mathbf{x}_p+\delta\mathbf{x}_c),
\qquad
\delta\mathbf{x}_c\in\left\{\pm h\,\Delta x\,\mathbf{e}_x \pm h\,\Delta x\,\mathbf{e}_y\right\},
\tag{4}
\end{equation}
where $N_i$ is the quadratic B-spline shape function and $h=\frac{1}{2\,\mathrm{PPC}}$ places the corner samples at a distance $\Delta x/(2\,\mathrm{PPC})$ from the particle center. For gradient-based quantities (e.g., velocity gradients used by artificial viscosity), we apply the same averaging to $\nabla N_i$.Unless explicitly noted, all subsequent experiments employ CPDI-lite transfers to reduce quadrature bias under imperfect particle sampling and to alleviate non-physical spike-head dents.

\paragraph{Particle-to-grid (P2G) mass and momentum}
Each material point carries mass $m_p$ and velocity $\mathbf{v}_p$. In APIC, an additional affine matrix $\mathbf{C}_p$ (a local velocity-gradient proxy) is stored to support angular-momentum-consistent transfers. The P2G mapping reads
\begin{equation}
m_i = \sum_p w_{ip} m_p, \qquad (m\mathbf{v})_i = \sum_p w_{ip} m_p\left(\mathbf{v}_p + \omega_p\,\mathbf{C}_p(\mathbf{x}_i-\mathbf{x}_p)\right). \tag{5}
\end{equation}
Here $\omega_p\in[0,1]$ is the soft-switch weight (Sec.~\ref{sec:undersampling_controls}). When $\omega_p=1$, Eq. (5) reduces to standard APIC; when $\omega_p=0$, only the translational contribution remains and the mapping is locally PIC-like. After accumulation, grid velocities are computed as $\mathbf{v}_i=(m\mathbf{v})_i/m_i$ and boundary conditions are enforced.

\paragraph{Grid-to-particle (G2P) update and FLIP blending}
After the grid momentum equation is advanced, material-point velocities are updated using a PIC/FLIP blend. Define the PIC velocity interpolation and FLIP increment as
\begin{equation}
\mathbf{v}_p^{\mathrm{PIC},n+1}=\sum_i w_{ip}\,\mathbf{v}_i^{n+1}, \qquad \Delta\mathbf{v}_p^{\mathrm{FLIP},n+1}=\sum_i w_{ip}\,(\mathbf{v}_i^{n+1}-\mathbf{v}_i^{n}).
\end{equation}
The blended update is
\begin{equation}
\mathbf{v}_p^{n+1}=(1-\alpha)\,\mathbf{v}_p^{\mathrm{PIC},n+1}+\alpha\,(\mathbf{v}_p^{n}+\Delta\mathbf{v}_p^{\mathrm{FLIP},n+1}). \tag{6}
\end{equation}
The parameter $\alpha\in[0,1]$ controls numerical dissipation: smaller $\alpha$ increases PIC-like damping, while larger $\alpha$ retains FLIP-like low dissipation. In APIC mode, $\mathbf{C}_p$ is reconstructed from the updated grid velocity field and used in the subsequent P2G step.

\paragraph{APIC affine reconstruction (implementation detail).}
For reproducibility, we use the standard APIC/MLS-MPM-style moment-based reconstruction of the affine matrix from
grid velocities. Define
\begin{equation}
\mathbf{D}_p = \sum_i w_{ip}\,(\mathbf{x}_i-\mathbf{x}_p)(\mathbf{x}_i-\mathbf{x}_p)^{T},
\qquad
\mathbf{B}_p = \sum_i w_{ip}\,\mathbf{v}_i^{n+1}(\mathbf{x}_i-\mathbf{x}_p)^{T},
\label{eq:apic_moments}
\end{equation}
and set
\begin{equation}
\mathbf{C}_p^{n+1} = \mathbf{B}_p\,(\mathbf{D}_p+\epsilon_D \mathbf{I})^{-1},
\label{eq:apic_recon}
\end{equation}
with a small $\epsilon_D$ to regularize near-singular configurations. When the soft-switch is enabled, we store
$\mathbf{C}_p^{n+1}\gets \omega_p\,\mathbf{C}_p^{n+1}$, matching Algorithm~\ref{algo:apicflip}.

\subsection{Initialization}
\label{sec:init_jitter}

CPDI-lite reduces quadrature bias for a given particle set, but it cannot compensate for systematic aliasing introduced by strictly lattice-aligned seeding. In particle--grid methods (PIC/FLIP/MPM/APIC), if particles are initialized on a perfectly regular subcell lattice, the particle phase relative to the grid interpolation locations repeats deterministically, which can lead to grid imprinting (phase locking) and long-time accumulation of non-physical streaks or void-like patterns. To decorrelate this error while preserving the macroscopic initial condition, we apply a small subcell perturbation to the particle position within each cell at $t=0$.~\cite{Steffen2008Quadrature,HammerquistNairn2017Noise}

Let $\boldsymbol{\xi}\in[0,1)^2$ denote the normalized local coordinates of a particle within its host cell. We perturb the phase by a pseudo-random offset $\delta\boldsymbol{\xi}$ and wrap it back into the unit cell:
\begin{equation}
\boldsymbol{\xi}' = (\boldsymbol{\xi} + \delta\boldsymbol{\xi}) \bmod 1,
\qquad
\|\delta\boldsymbol{\xi}\| = \mathcal{O}\!\left(\mathrm{PPC}^{-2}\right).
\tag{12}
\end{equation}
The perturbation amplitude is far below the cell scale, so it does not change the prescribed interface perturbation or macroscopic fields, but it breaks strict particle--grid alignment and noticeably improves numerical robustness in long-time RTI runs. Throughout the results in Figs.~3--10, we employ the widely used CPDI-lite transfer together with subcell-jittered particle seeding to mitigate quadrature bias and suppress grid imprinting.

We use an explicit time step restricted by a CFL condition based on the local acoustic speed. With ideal-gas closure $p=(\gamma-1)\rho e$, the sound speed is $c=\sqrt{\gamma p/\rho}$. The time step is chosen as
\begin{equation}
\Delta t = \mathrm{CFL}\,\frac{\Delta x}{\max_p (\|\mathbf{u}_p\|+c_p)}. \tag{7}
\end{equation}

We update internal energy using a work-based form consistent with the particle–grid velocity field and the pressure stress. Specifically, using the particle velocity gradient reconstructed from the grid,
\begin{equation}
\nabla \mathbf{u}(\mathbf{x}_p)=\sum_i \mathbf{v}_i \otimes \nabla w_{ip},
\end{equation}
the internal-energy update is advanced explicitly. Density is obtained from particle mass and volume, and pressure is computed using the EOS $p=(\gamma-1)\rho e$. (In implementation, a small pressure/energy floor may be applied to avoid negative pressures in extreme rarefactions; unless otherwise noted, the same floor is used for all methods compared.)

To stabilize shocks while preserving vortical structures, we adopt the vorticity-aware tensor artificial viscosity (AV) following Peddavarapu and Huang (2025). The tensor AV augments the pressure stress with an additional dissipative stress that activates primarily in compression and is reduced in strongly vortical regions, mitigating excessive dissipation in shear layers while maintaining robust shock capturing.

\subsection{Under-sampling controls: conservative resampling and soft-switch}
\label{sec:undersampling_controls}

\begin{figure}
    \centering
    \includegraphics[width=0.95\linewidth]{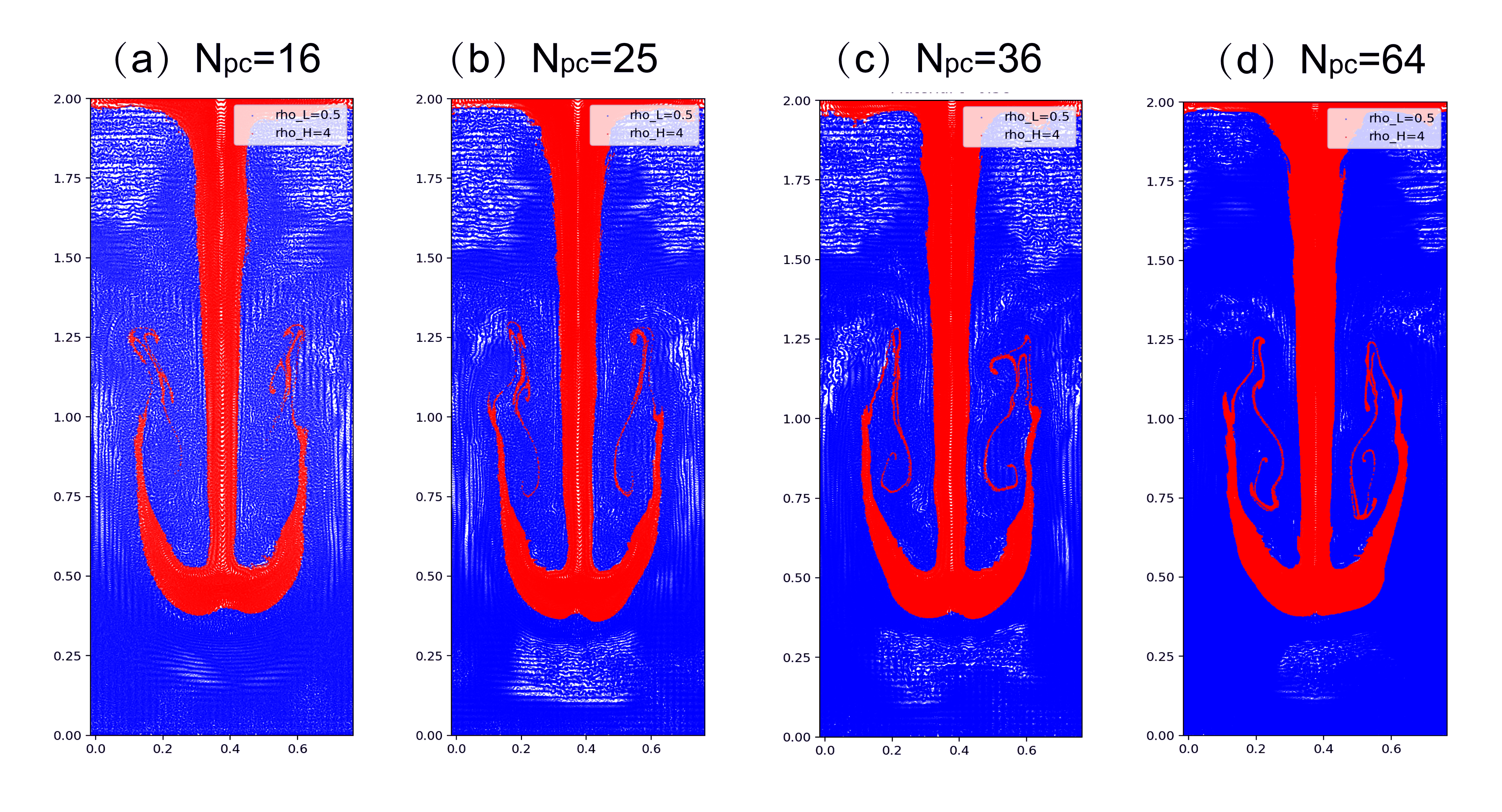}
    \caption{Single-mode RTI at $t=0.58$ with varying particles per cell (PPC, denoted $N_{pc}$). Increasing PPC alleviates the late-time spike-head ``dent'', indicating a sampling-driven origin.}
    \label{fig:ppc_sweep_dent}
\end{figure}

\begin{figure}
    \centering
    \includegraphics[width=0.95\linewidth]{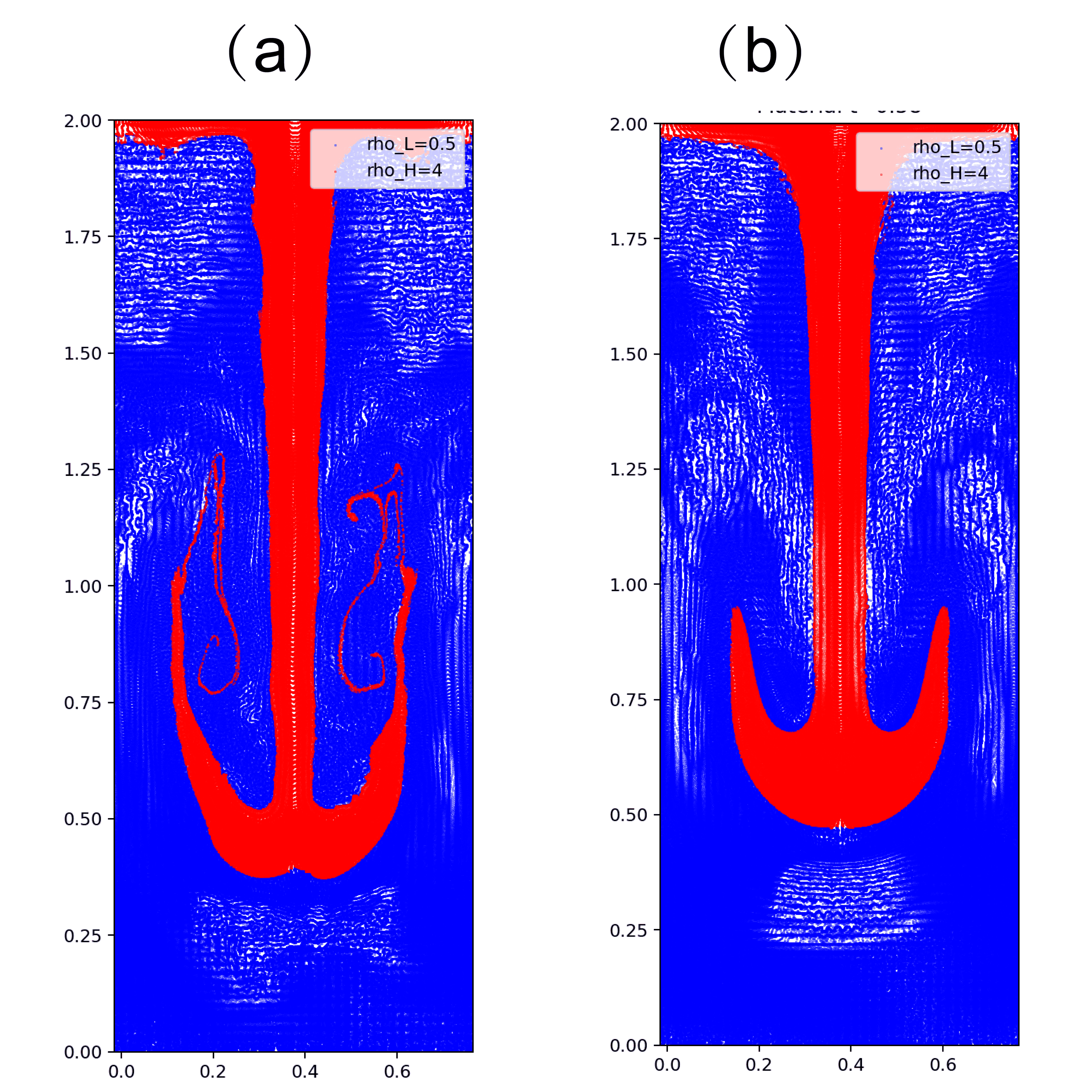}
    \caption{Single-mode RTI at $t=0.58$ with $N_{pc}=36$: (a) 98\% FLIP + 2\% APIC; (b) 98\% FLIP + 2\% PIC. Replacing the APIC affine transfer by PIC removes the dent but noticeably damps fine-scale roll-up.}
    \label{fig:picflip_ppc36}
\end{figure}

Long-time Rayleigh--Taylor instability (RTI) produces extreme material stretching and particle
redistribution near spike heads. As the interface rolls up, particles can locally evacuate the spike tip even
when the global particles-per-cell (PPC) is moderate, leading to two coupled numerical issues: (i) degraded
particle quadrature (insufficient support for particle--grid transfers) and (ii) unstable reconstruction of the
APIC affine matrix $\mathbf{C}_p$, whose contribution can inject spurious kinetic energy when local sampling is
poor. In our baseline FLIP--APIC + tensor-AV solver, these effects appear as a nonphysical, persistent
spike-head ``dent'' at late times.

We first establish that the dent is driven by under-sampling. Figure~\ref{fig:ppc_sweep_dent} shows a PPC sweep
at $t=0.58$ (with CPDI-lite transfers and subcell-jittered seeding enabled, unless otherwise stated): at low PPC,
the spike head exhibits a pronounced void-like indentation, whereas increasing PPC progressively restores the
spike-tip morphology. The reduction is monotone across the tested PPC values, indicating that the artifact is
tightly correlated with local particle support rather than with the global flow evolution. In particular, the
fact that the dent weakens as more particles are available per cell supports the interpretation that the spike
head becomes transiently particle-depleted in long-time runs, leading to quadrature deficiency and biased
particle--grid coupling.

We then isolate which part of the transfer is responsible for the remaining dents once sampling is improved.
In Fig.~\ref{fig:picflip_ppc36} (at $N_{pc}=36$ and the same physical time), the only change between the two runs
is whether the affine APIC term is included in the transfer. With 98\% FLIP + 2\% APIC, a residual dent persists
near the spike head, consistent with an ill-conditioned affine reconstruction under local depletion. Replacing
the APIC affine transfer by a PIC transfer (98\% FLIP + 2\% PIC) removes the dent, but the shear-layer roll-up
becomes visibly more diffuse: small-scale vortical structures are damped and the mushroom-cap features lose
finer details. Together, Figs.~\ref{fig:ppc_sweep_dent}--\ref{fig:picflip_ppc36} indicate that the issue is not
the base FLIP/PIC mixture itself, but the APIC affine component becoming unreliable when local sampling
deteriorates. This motivates a localized and minimally dissipative remedy with two complementary parts:
(i) restore particle support only where depletion occurs, and (ii) \emph{temporarily} gate the APIC affine term
only in those under-sampled regions. PIC is used here as a stabilizing limiter: although more diffusive, it
suppresses grid-scale noise and prevents an ill-conditioned affine estimate from injecting non-physical
interface artifacts.

\paragraph{(1) Conservative split resampling (restore local support).}
Let $n_c$ denote the particle count in cell $c$, and let $n_0$ be a target support level. When $n_c$ falls below
a threshold, we replenish quadrature points by conservative splitting. A parent particle $p$ is replaced by two
children $p_1,p_2$ that exactly conserve mass, momentum, and internal energy:
\begin{equation}
m_{p_k}=\tfrac12 m_p,\qquad \mathbf{v}_{p_k}=\mathbf{v}_p,\qquad e_{p_k}=e_p,\qquad k\in\{1,2\},
\label{eq:split_conservative_44}
\end{equation}
with positions $\mathbf{x}_{p_k}=\mathbf{x}_p\pm\delta\mathbf{x}$ using a small, cell-local jitter
$\delta\mathbf{x}$ to avoid coincident children. This directly increases $n_c$ (hence quadrature quality) in
depleted regions while leaving conserved quantities unchanged.

\paragraph{(2) Soft-switch on the APIC affine term (robust transfer under depletion).}
Even with resampling, long-time RTI can exhibit \emph{transient} depletion events where $\mathbf{C}_p$ is
ill-conditioned for a few steps. To prevent the affine mode from amplifying inconsistencies, we gate only the
affine contribution in the P2G transfer:
\begin{equation}
(m\mathbf{v})_i \;\; \mathrel{+}= \;\; w_{ip} m_p\Big[\mathbf{v}_p + \omega_p\,\mathbf{C}_p(\mathbf{x}_i-\mathbf{x}_p)\Big].
\label{eq:gated_apic_44}
\end{equation}
Thus $\omega_p=1$ recovers standard APIC, while $\omega_p=0$ disables the affine term (PIC-like transfer),
retaining FLIP blending for the translational component.

We choose $\omega_p$ as a product of a support-based factor and a flow-feature factor,
\begin{equation}
\omega_p = w_{\mathrm{Sup},p}\, w_{A\Omega,p},
\label{eq:omega_softswitch}
\end{equation}
so APIC is preserved in well-sampled, vortical regions and suppressed in poorly sampled or strongly compressive
regions.

\emph{Support weight.} Let $r_c=n_c/n_0$ be the support ratio for the cell containing particle $p$, and choose
thresholds $0<r_{\mathrm{low}}<r_{\mathrm{high}}$. We set
\begin{equation}
s_c=\mathrm{clamp}\!\left(\frac{r_c-r_{\mathrm{low}}}{r_{\mathrm{high}}-r_{\mathrm{low}}},\,0,\,1\right),\qquad
w_{\mathrm{Sup},p}=s_c^2(3-2s_c),
\label{eq:wsup_44}
\end{equation}
so that $w_{\mathrm{Sup},p}\approx 0$ in severely depleted cells and $w_{\mathrm{Sup},p}\approx 1$ in
well-supported regions. Conservative splitting increases $n_c$ and therefore drives $w_{\mathrm{Sup},p}$ back
toward $1$ between resampling steps.

\emph{Compression--vorticity weight.} Define the compression rate and vorticity magnitude at particles as
\begin{equation}
\chi_p=\max\!\left(0,-\nabla\!\cdot\mathbf{v}(\mathbf{x}_p)\right),
\qquad
\Omega_p=\left|\omega(\mathbf{x}_p)\right|,
\label{eq:chi_omega_def}
\end{equation}
where $\omega=\partial_x v_y-\partial_y v_x$ in 2D. At each step we normalize both to $[0,1]$ using
instantaneous maxima,
\begin{equation}
\tilde{\chi}_p=\frac{\chi_p}{\chi_{\max}+\varepsilon},\qquad
\tilde{\Omega}_p=\frac{\Omega_p}{\Omega_{\max}+\varepsilon'},
\label{eq:chi_omega_norm}
\end{equation}
and set
\begin{equation}
w_{A\Omega,p}=\frac{\tilde{\Omega}_p+\varepsilon}{\tilde{\chi}_p+\tilde{\Omega}_p+\varepsilon}\in[0,1],
\label{eq:wAomega_44}
\end{equation}
so that the affine term is suppressed in strongly compressive/weakly vortical regions but retained in
vortex-dominated shear layers.

In practice, we evaluate $\omega_p$ at the end of step $n$ and reuse it for the P2G transfer at step $n\!+\!1$
(lagged evaluation). This avoids extra passes over particles while providing a smooth APIC$\rightarrow$PIC
transition precisely where local support is inadequate.

\subsection{Ablation study (method rationale)}
\label{sec:ablation_method}

To justify the two under-sampling controls, we perform a component-wise study on single-mode RTI that isolates the
effects of (i) conservative resampling and (ii) the affine soft-switch. Starting from the prior-work baseline
FLIP--APIC with tensor artificial viscosity (tensor AV), we consider four variants:

\begin{itemize}
  \item (a) baseline: FLIP--APIC + tensor AV;
  \item (b) baseline + CPDI-lite;
  \item (c) baseline + CPDI-lite + conservative split resampling;
  \item (d) baseline + CPDI-lite + affine soft-switch;
  \item (e) full method: baseline + CPDI-lite + split resampling + soft-switch.
\end{itemize}

The comparison focuses on two diagnostics aligned with the failure analysis in
Sec.~\ref{sec:undersampling_controls}: (i) the absence of spike-head particle voids/dents in long-time evolution
and (ii) the preservation of physically plausible vortex roll-up without excessive smearing. As shown in
Fig.~5(a), the baseline configuration exhibits pronounced particle voids around the spike head together with a
clear nonphysical indentation. Enabling CPDI-lite together with subcell-jittered seeding (b) substantially
reduces the void-like sampling gaps and noticeably weakens the dent, consistent with reduced quadrature bias and
suppressed grid imprinting. Adding conservative split resampling on top of CPDI-lite and jitter (c) produces
only a marginal additional change in the dent for this case, indicating that splitting alone does not fully
address the residual spike-head indentation. This suggests that, in this regime, the remaining dent is dominated
by affine-mode instability rather than by a persistent lack of particle count. In contrast, combining CPDI-lite
and jitter with the affine soft-switch (d) further attenuates the dent by suppressing the APIC affine
contribution precisely when local particle support becomes transiently insufficient. Finally, the full method
(e), which couples CPDI-lite, conservative resampling, and the soft-switch (with jitter), eliminates the
spike-head dent in our tests while retaining fine-scale roll-up structures.

\begin{figure}
    \centering
    \includegraphics[width=0.95\linewidth]{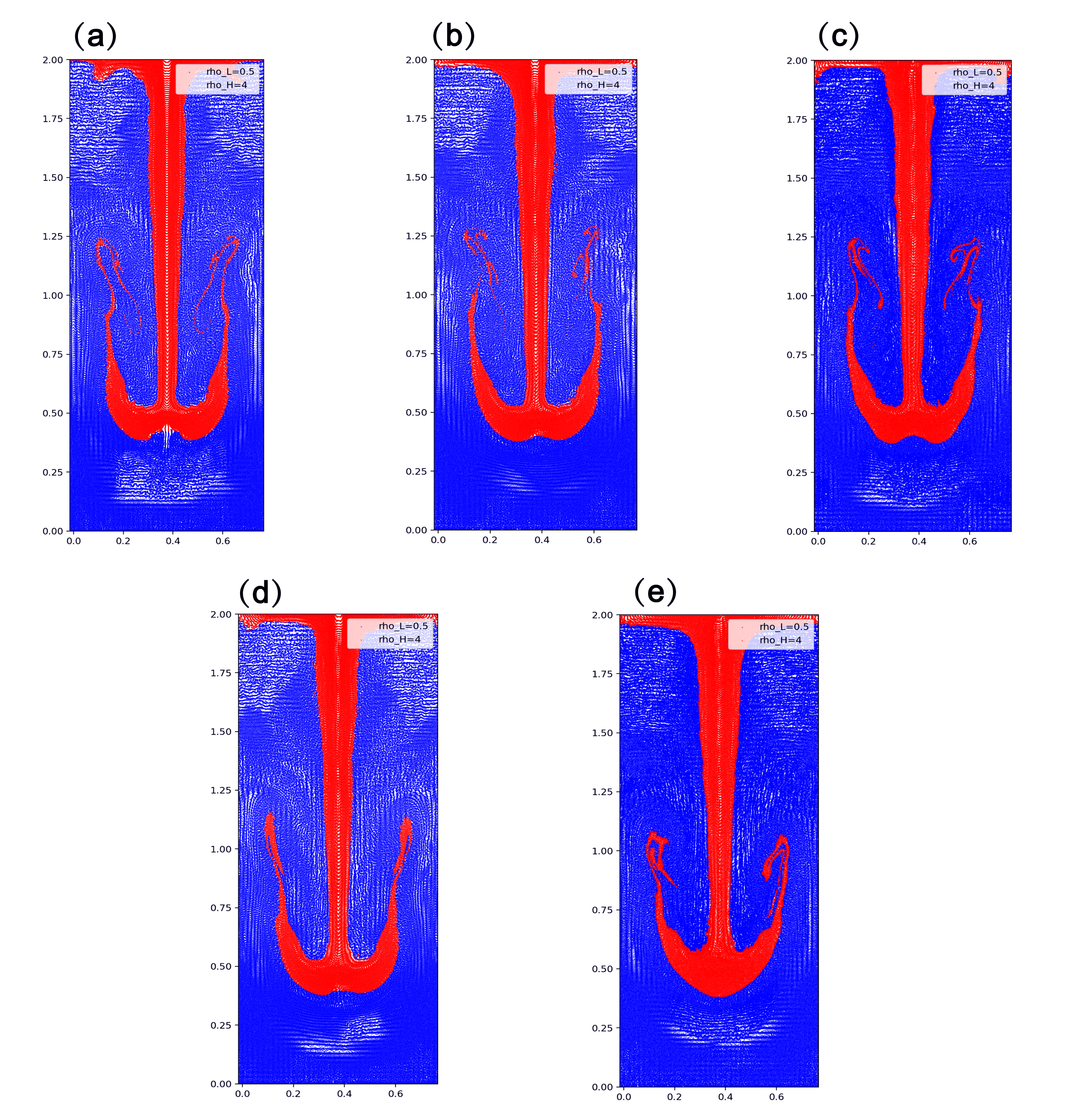}
    \caption{Ablation variants for single-mode RTI (representative late-time snapshot): (a) baseline: FLIP--APIC + tensor AV; (b) baseline + CPDI-lite (c) baseline + conservative split resampling; (d) baseline + CPDI-lite + affine soft-switch; (e) full method (baseline + CPDI-lite + split + soft-switch).}
    \label{fig:ablation_rti}
\end{figure}


\section{Numerical Experiments}

We assess accuracy and robustness using (A) the one-dimensional Sod shock tube, and (B) single-mode RTI.
Unless stated otherwise, we use quadratic B-splines for interpolation, CFL-based time stepping, and periodic
boundaries in $x$ for RTI.
\subsection{Implementation details and parameter choices}

\noindent\textbf{Software framework.} Our simulator is implemented in \emph{Taichi}, a Python-embedded domain-specific language for portable, high-performance parallel computing. Taichi uses a just-in-time (JIT) compiler to translate \emph{kernels} operating on data containers (fields) into optimized parallel code targeting multi-core CPUs and GPUs, while following a data-oriented programming model that decouples computation from data layout. These features make it well suited for particle--grid methods, where particle updates and particle--grid transfers can be expressed as massively parallel kernels with minimal low-level parallel programming overhead.\cite{Hu2019Taichi,Hu2020DiffTaichi}

We use a structured Cartesian background grid with quadratic B-spline shape functions. The time step is chosen to satisfy a CFL-like constraint based on the local acoustic speed and the maximum particle velocity. To avoid negative pressures in extreme rarefactions, a small pressure/energy floor may be applied in implementation; unless otherwise noted, reported results use the same floor across all methods compared.

Biindicator gating is applied to the affine (APIC) contribution only: the particle translational velocity is always transferred, while the affine term is multiplied by the gate \,(\(\omega_p\)). In practice the gate is computed from reconstructed velocity gradients and is therefore applied in a lagged fashion: \,(\(\omega_p^{n}\)) is computed at the end of step \((n)\) and used during the subsequent P2G. This avoids extra passes over particles and matches the implementation used in our experiments.

Unless stated otherwise, we use the default artificial-viscosity coefficients \((c_l,c_q)=(1,2)\) and A-FLIP blending \(\alpha=0.98\). The soft-switch thresholds are selected so that fully-resolved regions remain near \(\omega_p\approx1\), while strongly under-supported or strongly compressive regions smoothly transition toward PIC-like behavior (\(\omega_p\to0\)). Resampling is performed every \(N_{res}\) steps; in depleted cells we split particles to restore target support, and in over-dense cells we optionally merge to control cost while preserving conserved quantities.

\subsection{Sod shock tube}
We simulate the one-dimensional Sod shock tube problem~\cite{Sod1978} to benchmark shock capturing, contact
discontinuity resolution, and rarefaction propagation in a compressible MPM setting~\cite{DhakalZhang2016,Sulsky1994MPM,Stefan2023StrongShock}. The domain is
$x\in[0,1]$ with an initial discontinuity at $x_0=0.5$. The left and right states are
\begin{equation}
(u,\rho,p)\big|_{t=0}=
\begin{cases}
(0.0,\;1.0,\;1.0), & 0\le x < 0.5,\\
(0.0,\;0.125,\;0.1), & 0.5 < x \le 1.0,
\end{cases}
\label{eq:sod_ic}
\end{equation}
where the discontinuity is slightly smoothed following~\cite{Stefan2023StrongShock} to better isolate the scheme behavior and avoid
wall-heating effects. The background grid uses 400 cells, and we seed three material points per cell (PPC=3) to
ensure adequate quadrature. The simulation is advanced to $T=0.143$ with a fixed time step $\Delta t=10^{-4}$.

Figure~6 compares density, velocity, and pressure profiles at time $T$ with the exact Riemann solution. In all
cases, the proposed method captures the shock, contact discontinuity, and rarefaction with good agreement and
without spurious oscillations. Overall, the close match to the exact Riemann solution demonstrates that the tensor-AV-stabilized FLIP--APIC
framework achieves non-oscillatory shock capturing while maintaining accurate contact and rarefaction
resolution, establishing a robust compressible baseline for the subsequent long-time instability tests.

\begin{figure}
    \centering
    \includegraphics[width=0.95\linewidth]{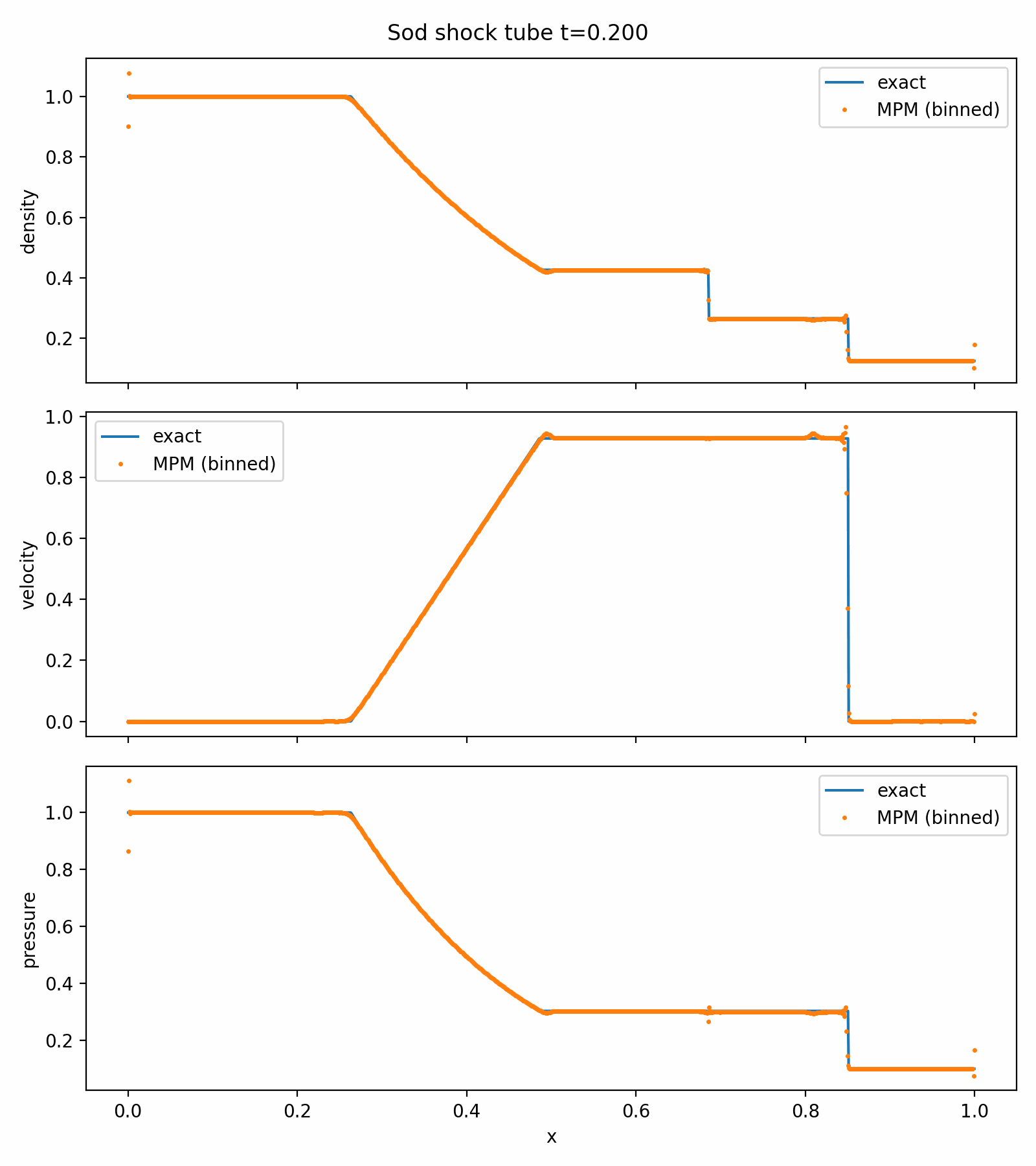}
    \caption{Sod shock tube at $t = 0.200$: density, velocity, and pressure profiles comparing the
proposed method (yellow line) against the exact Riemann solution (blue line).}
    \label{fig:ablation_rti}
\end{figure}

\subsection{Single-mode Rayleigh--Taylor instability}
We consider single-mode RTI in a rectangular domain $x\in[0,L_x]$, $y\in[0,L_y]$ with $y$ upward and a constant
downward body force $-g\,\mathbf{e}_y$. Unless otherwise stated, we use $L_x=0.75$ and $L_y=2.0$, periodic
boundary conditions in $x$, and slip/reflecting boundaries in $y$ (zero normal velocity at $y=0$ and $y=L_y$).
The heavy and light fluids have densities $\rho_H=4.0$ and $\rho_L=0.5$, respectively, with the heavy fluid
initialized above the interface and the light fluid below. The initial interface is prescribed by a single-mode
perturbation
\begin{equation}
y_{\mathrm{int}}(x)= y_0 + a\cos\!\left(\frac{2\pi m x}{L_x}\right),
\end{equation}
where $y_0$ is the mean interface height, $a$ is the amplitude, and $m$ is the mode number. We set $y_0=1.5$,
$a=0.1$, and $m=1$. The initial velocity is $\mathbf{v}(\mathbf{x},0)=\mathbf{0}$. We adopt an ideal-gas EOS with
$\gamma=1.4$. If desired, the material indicator can be regularized by a thin smoothing band of thickness
$\varepsilon$ across the interface (with $\varepsilon=0$ recovering a sharp interface).

\paragraph{Hydrostatic initialization and compressibility level.}
To avoid spurious acoustic transients at $t=0$, we initialize the pressure in hydrostatic equilibrium for a
\emph{flat} interface at $y=y_0$ using a piecewise-constant density profile,
\begin{equation}
p_0(y)=p_{\mathrm{ref}}-\rho_{\mathrm{flat}}(y)\,g\,(y-y_0),\qquad
\rho_{\mathrm{flat}}(y)=
\begin{cases}
\rho_H, & y>y_0,\\
\rho_L, & y\le y_0,
\end{cases}
\label{eq:hydrostatic_init}
\end{equation}
with a reference pressure $p_{\mathrm{ref}}=10$ specified at $y=y_0$. The particle internal energy is then set
from the EOS using the \emph{actual} density assigned by the perturbed interface,
\begin{equation}
e_0(\mathbf{x})=\frac{p_0(y)}{(\gamma-1)\rho_{\mathrm{eff}}(\mathbf{x})},
\qquad
\rho_{\mathrm{eff}}(\mathbf{x})=
\begin{cases}
\rho_H, & y>y_{\mathrm{int}}(x),\\
\rho_L, & y\le y_{\mathrm{int}}(x),
\end{cases}
\label{eq:e0_init}
\end{equation}
(with a smooth transition if interface smoothing is enabled). The choice of $p_{\mathrm{ref}}$ controls the
compressibility level: since the sound speed scales as $c\sim\sqrt{\gamma p_{\mathrm{ref}}/\rho}$, a lower
$p_{\mathrm{ref}}$ yields a smaller acoustic speed and thus a larger effective Mach number for the same
gravity-driven velocity scale. Using $U\sim\sqrt{gL_y}$, we obtain $U\approx\sqrt{15\times 2}\approx 5.48$ and
$c_H\approx\sqrt{\gamma p_{\mathrm{ref}}/\rho_H}\approx\sqrt{1.4\times 10/4}\approx 1.87$, giving
$\mathrm{Ma}_H\approx U/c_H\approx 2.9$ (and $\mathrm{Ma}_L\approx 1.0$ for the light fluid). This places the
setup in a strongly compressible RTI regime.

The domain is discretized by a uniform background grid with $n_x\times n_y$ cells and particles seeded at
$\mathrm{PPC}=4\times 4$ per cell.

Figure~7 compares single-mode RTI density (or material-indicator) snapshots between our MPM results and a
reference Euler solver at matched physical times, illustrating bubble rise and spike descent. In addition to the
visual comparison, we report the spike-tip position as a function of time to quantify the instability
development.

For the Euler solver, we use the open-source \texttt{pyro2} codebase (\url{https://github.com/python-hydro/pyro2})
with the same physical parameters as in our MPM runs. The Euler simulation is performed on a $150\times 400$ grid,
which is  $4\times$ higher resolution than the baseline MPM grid used for particle sampling. Despite
this higher resolution, the Euler solution exhibits noticeably stronger numerical smearing: the interface appears
more diffuse and small-scale flow structures are less sharply resolved. In contrast, our MPM results maintain a
sharper material interface with reduced non-physical mixing, yielding cleaner spike-head morphology under strong
stretching.

To provide a quantitative check, we track the spike-tip position $y_{\mathrm{sp}}(t)$ over time. The spike tip is
extracted consistently at each output time (using the material indicator/density to identify the heavy-fluid
interface and recording the leading spike location). Figure~8 shows that the spike-tip trajectory from our MPM
method agrees closely with the Euler reference in the overall growth and nonlinear evolution.

\begin{figure}
    \centering
    \includegraphics[height=0.96\textheight,keepaspectratio]{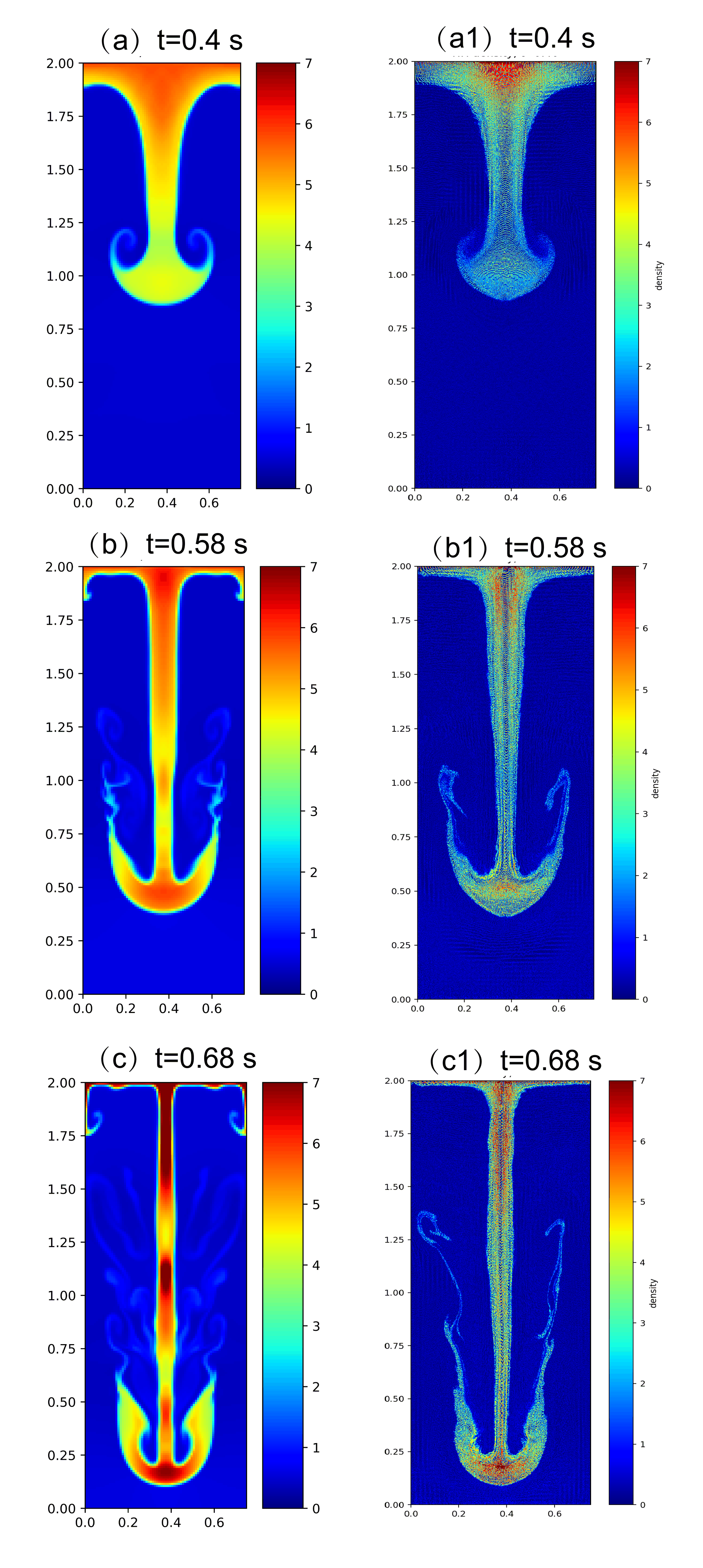}
    \caption{Single-mode RTI density/material-indicator snapshots at representative times, comparing the proposed
MPM method with the \texttt{pyro2} Euler reference (Euler grid resolution $150\times 400$).
\textbf{Left column:} Euler (\texttt{pyro2}); \textbf{right column:} MPM. }
    \label{fig:ablation_rti}
\end{figure}

\begin{figure}
    \centering
    \includegraphics[width=0.95\linewidth]{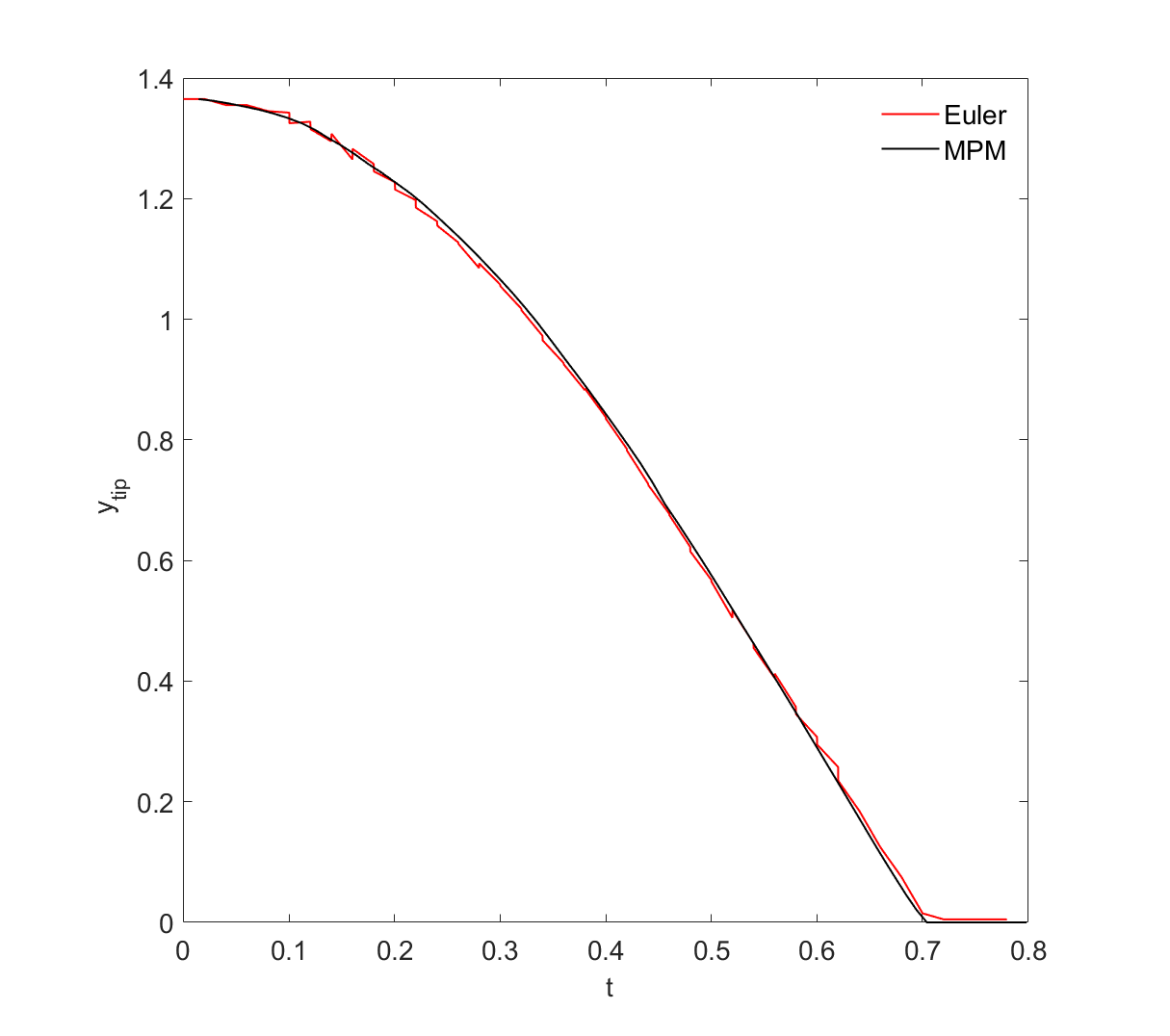}
    \caption{Single-mode RTI quantitative comparison: spike-tip position $y_{\mathrm{sp}}(t)$ versus time
for the proposed MPM method and the \texttt{pyro2} Euler reference.}
    \label{fig:ablation_rti}
\end{figure}

\subsection{Multi-mode Rayleigh--Taylor instability}
Multi-mode RTI involves broadband interfacial perturbations and sustained mixing, and is therefore more
challenging for sampling-limited particle methods than the single-mode case. In addition to qualitative
morphology, we evaluate two diagnostics directly tied to the failure analysis: (i) the absence of spike-head
dents/void-like artifacts in long-time evolution, and (ii) the preservation of physically plausible,
vortex-rich roll-up and mixing-layer growth---i.e., avoiding both excessive numerical diffusion that erases
small scales and particle noise that seeds nonphysical features.

We test the proposed method on a multi-mode RTI by perturbing the initial interface with a superposition of
Fourier modes,
\begin{equation}
y(x,0) = y_0 + \sum_k a_k \cos\!\left(\frac{2\pi m_k x}{L_x} + \phi_k\right),
\end{equation}
where phases $\phi_k$ and amplitudes $a_k$ are chosen to form a broadband spectrum while keeping the overall
perturbation amplitude small. Unless noted otherwise, we use the same domain, density ratio, gravity, boundary
conditions, and numerical parameters as in Sec.~5.2. We report material-indicator/density snapshots to assess
morphology and roll-up, together with coarse quantitative measures such as spike/bubble heights and
mixing-layer growth.

Qualitatively, the prior-work baseline can remain stable at early times but may develop local particle depletion
near rapidly stretching spike tips during long-time evolution, which biases the local quadrature and can appear
as depressions/void-like artifacts. In contrast, the proposed strategy replenishes quadrature only where needed
and suppresses poorly supported affine modes only when local sampling becomes unreliable, yielding robust
long-time evolution without sacrificing vortex-rich mixing structures in well-sampled regions.

Figure~9 compares density snapshots at matched times (left column: Euler/\texttt{pyro2}; right column: MPM).
At $t=0.4$ (Fig.~9(a,a1)), both solvers show similar early-stage bubble rise and spike descent, although the
Euler interface is already slightly more smeared. By $t=0.58$ (Fig.~9(b,b1)), roll-up structures are visible in
both solutions; however, the Euler field exhibits a broader mixed region, whereas MPM maintains a sharper
material boundary (with expected sampling-level texture near the interface). At $t=0.68$ (Fig.~9(c,c1)), the
overall morphology remains consistent while the Euler solution appears more diffuse in the shear layer and near
spike heads, suggesting stronger numerical mixing despite the higher grid resolution.

To provide a quantitative check, Fig.~10 shows the spike-tip trajectory $y_{\mathrm{tip}}(t)$. The two methods
agree closely in the early and mid stages, indicating that the dominant instability growth is captured
consistently. A modest late-time deviation is observed, with MPM exhibiting slightly slower spike penetration.
We conjecture that this deviation is that  the more diffuse interface in the Euler solution,
which can alter the late-time buoyancy--drag balance. Despite this modest discrepancy, the overall spike growth
trend and morphology remain consistent between the two solvers.

We also briefly examine robustness to key numerical parameters, including (i) particle-per-cell (PPC)
resolution, (ii) A-FLIP blending $\alpha$, and (iii) the smoothstep thresholds used in the Biindicator gate. As
PPC decreases, the baseline becomes increasingly prone to quadrature-driven artifacts and localized void-like
dents near rapidly deforming interface features. The proposed resampling mitigates these artifacts by restoring
particle support, while the Biindicator gate reduces affine energy injection in regions flagged as
under-supported or strongly compressive. Increasing $\alpha$ tends to reduce diffusion but can amplify noise;
our default $\alpha=0.98$ provides a stable trade-off when combined with tensor AV and the gating strategy.
Finally, the method is insensitive to moderate changes in the smoothstep thresholds, provided the support gate
saturates to $\omega_p\approx 1$ in well-sampled regions and transitions smoothly (without hard clipping) in
depleted cells.

For the multi-mode case, spike-tip tracking is inherently noisier due to multiple competing spikes. We therefore
use the same definition as in the single-mode case but optionally apply mild temporal smoothing when reporting
the curve. Consistent spike growth relative to the reference solver, together with improved morphological
stability, supports the effectiveness of the combined resampling and gating strategy.

\begin{figure}
    \centering
    \includegraphics[height=0.96\textheight,keepaspectratio]{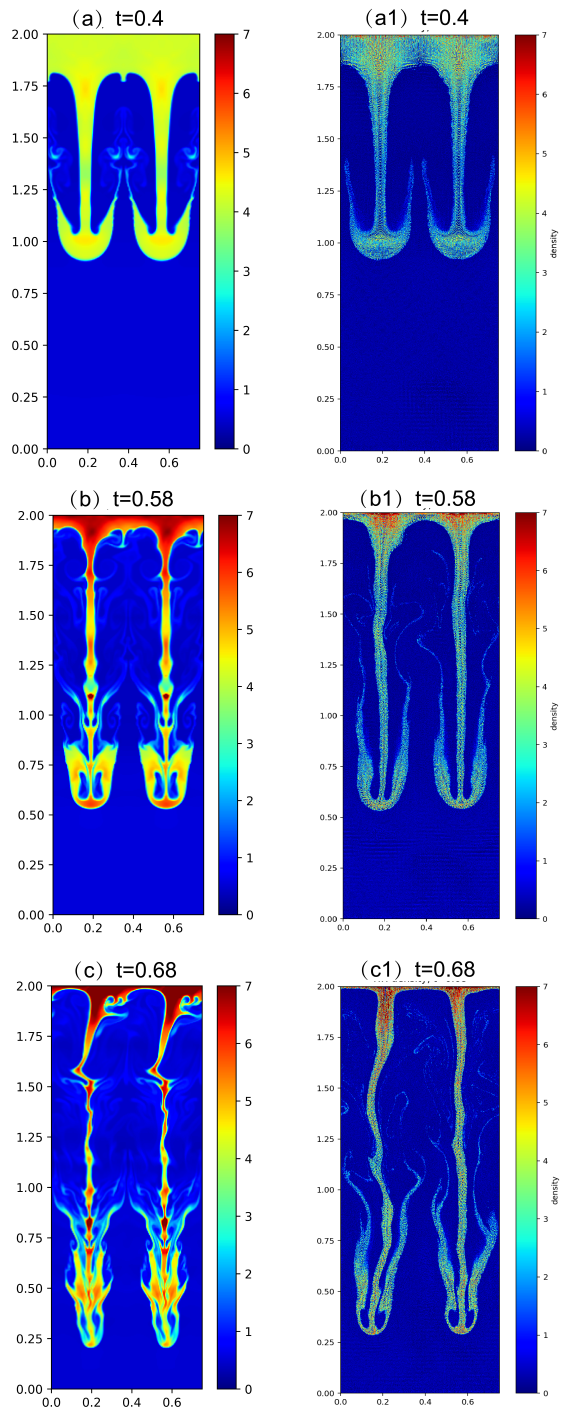}
    \caption{Single-mode RTI density snapshots at representative times ($t=0.4,\,0.58,\,0.68$).
\textbf{Left column:} reference Euler solution (\texttt{pyro2}, $150\times 400$ grid);
\textbf{right column:} proposed MPM results at the same times.
Both methods exhibit consistent bubble/spike morphology and roll-up development, while the Euler solution
appears more diffuse (stronger numerical mixing) and the MPM solution preserves a sharper interface with
sampling-level texture near the material boundary.}
    \label{fig:ablation_rti}
\end{figure}

\begin{figure}
    \centering
    \includegraphics[width=0.95\linewidth]{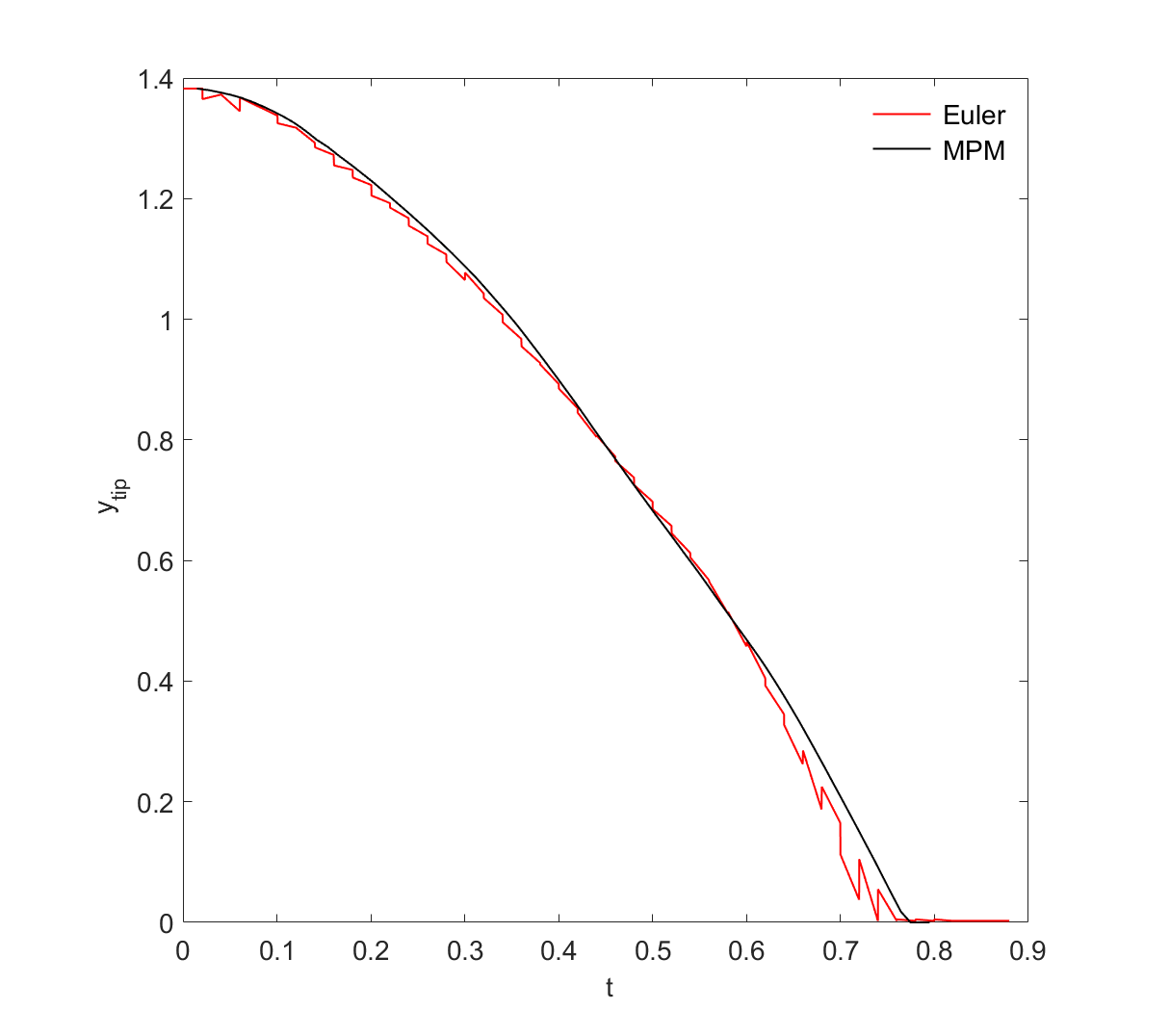}
    \caption{Multi-mode RTI quantitative comparison: spike-tip position $y_{\mathrm{sp}}(t)$ versus time
for the proposed method and the reference Euler solver pyro2.}
    \label{fig:ablation_rti}
\end{figure}

\section{Discussion and Conclusions}

The ablation study elucidates the roles of the proposed components beyond the FLIP--APIC + tensor artificial
viscosity (tensor-AV) baseline. While the baseline is robust for shock- and discontinuity-dominated flows and
reliably reproduces the Richtmyer--Meshkov instability, long-time Rayleigh--Taylor instability (RTI) remains
susceptible to particle depletion near spike heads. Such under-sampling degrades particle quadrature and
particle--grid coupling, which may manifest as a thermodynamic bias as well as nonphysical void-like
depressions in density and pressure.

Conservative split resampling directly targets this failure mode by replenishing quadrature points in depleted
cells while exactly preserving mass, momentum, and internal energy, thereby restoring local coupling quality.
However, resampling alone is not always sufficient: even with periodic resampling, extreme stretching can
transiently reduce particle support to an ill-conditioned state between resampling steps. To address this, the
proposed soft-switch (Biindicator gate) adaptively attenuates the APIC affine contribution in under-resolved
regions, preventing spurious kinetic-energy injection, while retaining full APIC fidelity in well-sampled shear
layers to preserve shear-driven vortex roll-up.

Several limitations warrant discussion. First, although the results demonstrate improved robustness across
representative shock and RTI benchmarks, a rigorous entropy-stability analysis is beyond the scope of the
present work. Second, the numerical parameters---including AV coefficients, resampling frequency, and
soft-switch thresholds---may require modest tuning across different problem classes, and an automatic
calibration strategy remains an open practical question. Third, the current study is restricted to
two-dimensional inviscid flows; extension to three dimensions is expected to introduce more complex depletion
patterns and higher computational costs, potentially necessitating adaptive resampling criteria together with
efficient parallel load balancing. Finally, while the conservative resampling strategy preserves mass,
momentum, and internal energy, multi-material configurations with distinct equations of state and explicit
interface tracking may require additional considerations. Moreover, conservative merge operations may introduce
mild local diffusion in energy and vorticity if applied aggressively, motivating careful criteria to avoid
over-coarsening in highly vortical mixing layers.

\paragraph{Key conclusions}
\begin{enumerate}
\item \textbf{Baseline limitation (diagnosis).}
The FLIP--APIC + tensor-AV baseline is robust for shock/discontinuity-dominated problems (e.g., RM), but long-time
RTI can become under-sampled near spike heads, leading to quadrature-induced thermodynamic bias and nonphysical
void-like depressions.

\item \textbf{Conservative resampling (quadrature restoration).}
Conservative split resampling replenishes particles in depleted cells while exactly conserving mass, momentum,
and internal energy, mitigating spike-head void formation by restoring local particle--grid coupling quality.

\item \textbf{Soft-switch (affine stabilization).}
An under-sampling-aware soft-switch is crucial for handling transient extreme depletion between resampling steps;
attenuating the APIC affine term in poorly supported regions suppresses spurious kinetic-energy injection while
preserving APIC accuracy in well-sampled shear layers.
\end{enumerate}

Together with vorticity-aware tensor AV, these mechanisms enable simultaneous shock robustness, suppression of
spike-head voids in long-time RTI, and preservation of shear-driven vortex roll-up. Future work will focus on:
(i) three-dimensional extensions with scalable neighborhood search and resampling; (ii) adaptive resampling and
gating criteria guided by error indicators, e.g., local quadrature residuals or entropy production; (iii)
higher-order or reproducing-kernel transfers in well-sampled regions while retaining robust gating in depleted
cells; and (iv) integration with adaptive grid refinement to reduce computational cost in smooth regions without
sacrificing shock resolution..


\end{document}